\documentclass[aps,prl,preprintnumbers,superscriptaddress,nofootinbib,twocolumn]{revtex4}
\usepackage[dvips]{graphicx}
\usepackage{bm,latexsym,amsmath,amssymb,amsfonts,mathrsfs}
\usepackage{color}
\input{colordvi.tex}
\newcommand*{\D}{{\rm d}}

\newcommand*{\mpl}{M_{\rm Pl}}
\begin{document}

\title{G-inflation: inflation driven by the Galileon field\footnote{This
paper is the original version of the article published in
Phys.\ Rev.\ Lett.\ {\bf 105}, 231302 (2010)
with the title ``Inflation driven by the Galileon field.''
}}

\author{Tsutomu~Kobayashi}
\affiliation{Research Center for the Early Universe (RESCEU), Graduate School of Science,
The University of Tokyo, Tokyo 113-0033, Japan}

\author{Masahide~Yamaguchi}
\affiliation{Department of Physics, Tokyo Institute of Technology, Tokyo 152-8551, Japan}

\author{Jun'ichi~Yokoyama}
\affiliation{Research Center for the Early Universe (RESCEU), Graduate School of Science,
The University of Tokyo, Tokyo 113-0033, Japan}
\affiliation{Institute for the Physics and Mathematics of the Universe(IPMU),
The University of Tokyo, Kashiwa, Chiba, 277-8568, Japan}

\begin{abstract}
We propose a new class of inflation model, {\em G-inflation}, which has a Galileon-like
nonlinear derivative interaction of the form $G(\phi, (\nabla\phi)^2)\Box\phi$
in the Lagrangian with the resultant equations
of motion being of second order. It
is shown that (almost) scale-invariant curvature fluctuations can be
generated even in the exactly de Sitter background and that the
tensor-to-scalar ratio can take a significantly larger value than in the
standard inflation models, violating the standard consistency
relation. Furthermore, violation of the null energy condition can
occur without any instabilities.  As a result, the spectral index of
tensor modes can be blue, which makes it easier to observe quantum
gravitational waves from inflation by the planned gravitational-wave
experiments such as LISA and DECIGO as well as by the upcoming CMB
experiments such as Planck and CMBpol.
\end{abstract}

\preprint{RESCEU-18/10}
\maketitle

Inflation in the early universe~\cite{inflation} is now a part of the
standard cosmology to solve the horizon and flatness problem as well as
to account for the origin of density/curvature fluctuations.  It is most
commonly driven by a scalar field dubbed as inflaton, and the research
on inflationary cosmology has long been focused on the shape of the
inflaton potential in the particle physics context.
Its underlying physics is now being probed using
precision observations of the cosmic microwave background~\cite{wmap}
and large scale structure which are sensitive only to the dynamical
nature of the inflaton.  Reflecting this situation, a number of novel
inflation models have been proposed extending the structure of the
kinetic function, such as k-inflation~\cite{kinflation}, ghost
condensate~\cite{ghost}, and DBI inflation~\cite{DBI}.

In this {\em Letter}, we propose a new class of inflation models, for
which the scalar field Lagrangian is of the form
\begin{eqnarray}
{\cal L}_\phi = K(\phi, X)-G(\phi, X)\Box\phi,\label{Galileon-Lagrangian}
\end{eqnarray}
where $K$ and $G$ are general function of $\phi$ and
$X:=-\nabla_\mu\phi\nabla^\mu\phi/2$.  The most striking property of
this generic Lagrangian~(\ref{Galileon-Lagrangian}) is that it gives
rise to derivatives no higher than two both in the gravitational- and
scalar-field equations.  In the simplest form the nonlinear term may be
given by $G\Box\phi\propto X\Box\phi$, which has recently been discussed
in the context of the so-called {\em Galileon} field~\cite{G1, G2}.
The general form $G(\phi, X)\Box\phi$ may be regarded as an extension of
the Galileon-type interaction $X\Box\phi$ 
while maintaining the field equations to be of second-order~\cite{vikman}.
So far the phenomenological aspects of the Galileon-type scalar field
have been studied mainly in the context of dark energy and modified
gravity~\cite{gde}. In this Letter, we discuss primordial inflation
induced by this type of fields.

Now let us start investigating our model in detail.  Assuming that
$\phi$ is minimally coupled to gravity, the total action
is given by
\begin{eqnarray}
S=\int\D^4x\sqrt{-g}\left[\frac{\mpl^2}{2}R+{\cal L}_\phi\right].\label{action}
\end{eqnarray}
The energy-momentum tensor $T_{\mu\nu}$
derived from the action reads
\begin{eqnarray}
T_{\mu\nu}&=&K_{X}\nabla_\mu\phi\nabla_\nu\phi+K g_{\mu\nu}-2\nabla_{(\mu}G \nabla_{\nu)}\phi
\nonumber\\&&
+ g_{\mu\nu}\nabla_{\lambda }G\nabla^{\lambda }\phi
-G_X\Box\phi\nabla_\mu\phi\nabla_\nu\phi.\label{tmn}
\end{eqnarray}
The equation of motion of the scalar field is equivalent to $\nabla_\nu T_\mu^{\;\nu}=0$.
Here and hereafter we use the notation $K_X$ for $\partial K/\partial X$ etc.

Taking the homogeneous and isotropic background,
$\D s^2=-\D t^2+a^2(t)\D\mathbf{x}^2$, $\phi=\phi(t)$,
let us study inflation driven by
the Galileon-like scalar field~(\ref{Galileon-Lagrangian}),
which we call ``{\em G-inflation}.''
The energy-momentum tensor~(\ref{tmn}) has the form $T_\mu^{\;\nu}=\text{diag}(-\rho, p, p, p)$ with
\begin{eqnarray}
\rho&=&2K_XX-K+3G_XH\dot\phi^3-2G_\phi X,
\label{eqn:EM}
\\
p&=&K-2\left(G_\phi+G_X\ddot\phi \right)X.
\end{eqnarray}
Here, $\rho$ has an explicit dependence on the Hubble rate $H$.
The gravitational field equations are thus given by
\begin{eqnarray}
3\mpl^2H^2=\rho,
\quad
-\mpl^2\left(3H^2+2\dot H\right)=p,
\end{eqnarray}
and the scalar field equation of motion reads
\begin{eqnarray}
 && K_X \left( \ddot{\phi}+3H\dot{\phi} \right) + 2 K_{XX} X \ddot{\phi} +
   2K_{X\phi} X - K_{\phi} \nonumber \\
&& -2\left( G_{\phi}-G_{X\phi}X \right)
     \left( \ddot{\phi}+3H\dot{\phi} \right) 
     + 6G_{X} \left[ \left( HX \right)\dot{}+3H^2 X \right] \nonumber \\
&& - 4G_{X\phi} X \ddot{\phi} - 2G_{\phi\phi}X+6HG_{XX}X \dot{X}=0.
\label{eqn:EOM}
\end{eqnarray}
These three equations constitute two independent evolution equations for
the background. Note that the appearance of the terms proportional to
the Hubble parameter in Eqs. (\ref{eqn:EM}) and (\ref{eqn:EOM}) reflects
the fact that the Galileon symmetry is broken in the curved spacetime
even if we constrain our functional form of the Lagrangian which possess
its symmetry in the Minkowski spacetime.

We begin with constructing an exactly de Sitter background, 
%
taking $K$ and $G$ as
\begin{eqnarray}
K(\phi, X)=K(X),\quad G(\phi, X)=g(\phi)X.\label{Ginf}
\end{eqnarray}
In this case, inflation is driven purely kinematically, although G-inflation
does not preclude a potential-driven inflationary solution with an explicit $\phi$-dependence
in $K(\phi, X)$ in general; see Eq.~(\ref{Galileon-Lagrangian}).
If $g(\phi)=$ const,
{\em i.e.}, the Lagrangian has a shift symmetry $\phi\to\phi +$ const,
we have an exactly de Sitter solution satisfying
$\dot\phi=$const,
\begin{eqnarray}
3\mpl^2 H^2 = -K,
\quad
{\cal D}:=K_X+3gH\dot\phi=0.
\end{eqnarray}

Let us now provide a simple example:
\begin{eqnarray}
K=-X+\frac{X^2}{2M^3\mu},
\quad g=\frac{1}{M^3},
\label{toymodel}
\end{eqnarray}
where $M$ and $\mu$ are parameters having dimension of mass.
The de Sitter solution is given by
\begin{eqnarray}
X =M^3\mu x,\quad
H^2=\frac{M^3}{18\mu }\frac{(1-x)^2}{x},
\end{eqnarray}
where $x$ ($0<x<1$) is a constant satisfying
$(1-x)/x\sqrt{1-x/2} = \sqrt{6}\mu/\mpl$.
For $\mu\ll\mpl$, it can be seen that
$x\simeq 1-\sqrt{3} \mu/\mpl$ and hence
the Hubble rate during inflation is given in terms of
$M$ and $\mu$ as $H^2 \simeq M^3\mu/(6 \mpl^2)$.
As the first term in $K(X)$ has the ``wrong'' sign,
one may worry about ghost-like instabilities.
However, as we will see shortly, this model
is free from ghost and any other instabilities.

By allowing for a tilt of the function $g(\phi)$
we can find quasi-de Sitter inflation, which also induces a tilt in the
spectral index of curvature fluctuations; see Eq.~(\ref{sind}) below.
To investigate this possibility we define
\begin{eqnarray}
\epsilon := -\frac{\dot H}{H^2},
\quad
\eta := -\frac{\ddot\phi}{H\dot\phi},
\quad
\epsilon_g := \mpl \frac{g_\phi}{g}.
\end{eqnarray}
The slow-roll conditions are given by $|\epsilon| \ll 1$ and $|\eta|
\ll 1$, one of which can be replaced by $|\dot g/H g| \ll 1$ as long as
$X K_{XX}/K_X = {\cal O}(1)$. The slow-roll condition $|\dot g/H g| \ll 1$
can also be written in terms of $\epsilon_g$ as $|\epsilon_g|\sqrt{X/(-K)}\ll 1$.
In order to have $\rho\simeq -K$ we require $|X{\cal D}/K|\ll 1$.
Thus, the ``slow-roll'' equations are $3\mpl^2H^2\simeq -K$ and 
${\cal D}\simeq 0$. 
From the first equation and its time derivative, $6\mpl^2H\dot H\simeq -K_X\dot X$,
we find that $\epsilon$ and $\eta$ are actually related as
$\epsilon \simeq \eta XK_X/K.$ 

For a toy model with $\epsilon_g=$ const, namely,
$g(\phi)=e^{\epsilon_g \phi/\mpl}/M^3$,
and various $K(X)$, we have solved numerically the relevant equations,
and confirmed that the quasi-de Sitter solution is an attractor.

Inflation can be terminated by 
incorporating the $\phi$-dependence of the linear term in the kinetic function,
\begin{eqnarray}
K(\phi, X)=-A(\phi)X+\Delta K,
\end{eqnarray}
to flip the sign of $A$ to the ``normal'' one ($A=$ const $<0$)
due to the nontrivial evolution of $A(\phi)$ [$\simeq A(\dot\phi t)$]
in the final stage of inflation,
while $G(\phi, X)$ may still be of the form
$G=g(\phi)X$ with $g\simeq$ const.
As an explicit example, one can take
$K=-A(\phi)X+X^2/2M^3\mu$ and $G=X/M^3$
where $A(\phi)=\tanh[\lambda(\phi_{\rm end}-\phi)/\mpl]$ 
with $\lambda={\cal O}(1)$.
Our numerical solution shows that soon after $\phi$ crosses $\phi_{\rm end}$
to change the sign of $A$, it stalls and all the higher-order terms $\Delta K$
as well as terms from $G\Box\phi$ become negligibly small
within one e-fold.
As a result, $\phi$ behaves as a massless canonical field,
so that the energy density of the scalar field is diluted as rapidly as $\rho\propto a^{-6}$.

Since the shift symmetry of the original Lagrangian prevents
direct interaction between $\phi$ and standard-model contents,
reheating proceeds only through gravitational particle production
as discussed by Ford~\cite{ford},
who has shown that at the end of inflation ($a=1$) there exists
radiation with its energy density corresponding at least to the Hawking temperature:
\begin{eqnarray}
\left.\rho_r\right|_{\rm end}=\frac{\pi^2}{30}g_*T_H^4,\quad T_H=\frac{H_{\rm inf}}{2\pi}.
\end{eqnarray}
Since
$\rho_r=3\mpl^2H_{\rm inf}^2(g_*/1440\pi^2)(H_{\rm inf}/\mpl)^2 \ll 3\mpl^2H_{\rm inf}^2$,
the radiation component does not affect the cosmic history at that moment.
Subsequently, the radiation energy density decays as $\rho_r\propto a^{-4}$,
while
the energy density of $\phi$ is
diluted more rapidly as $\rho_\phi\propto a^{-6}$.
Finally one finds $\rho_\phi=\rho_r$
at $a=\sqrt{3}\mpl H_{\rm inf}\left(\left.\rho_r\right|_{\rm end}\right)^{-1/2}$.
Defining the reheating temperature by 
$\rho_r=\rho_\phi=(\pi^2/30)g_*'T_R^4$,
one can estimate
\begin{eqnarray}
T_R \simeq 0.01\,\frac{H^2_{\rm inf}}{\mpl} \simeq
     10^{4}\,{\rm GeV} \left(\frac{H_{\rm inf}}{10^{12}\,{\rm GeV}}\right)^2.
\end{eqnarray}

Let us briefly comment on baryogenesis through leptogenesis
\cite{Fukugita:1986hr}. If the mass of the lightest right handed
neutrinos $N_1$ is smaller than the Hawking temperature $T_H$, they are
copiously produced at the end of inflation. Then, their out-of
equilibrium decay can lead to lepton asymmetry, which is converted to
baryon asymmetry through the sphaleron effects. In our scenario, the
decay parameter, which is defined as the ratio of the decay rate of
$N_1$ to the Hubble parameter when $N_1$ becomes non-relativistic, is
significantly suppressed compared to the standard thermal leptogenesis
scenario \cite{Fukugita:1986hr} because the universe is dominated not by
radiations but by the inflaton. Thus, the wash out processes are
significantly suppressed and sufficient baryon asymmetry is produced.

\begin{figure}[tb]
  \begin{center}
    \includegraphics[keepaspectratio=true,height=55mm]{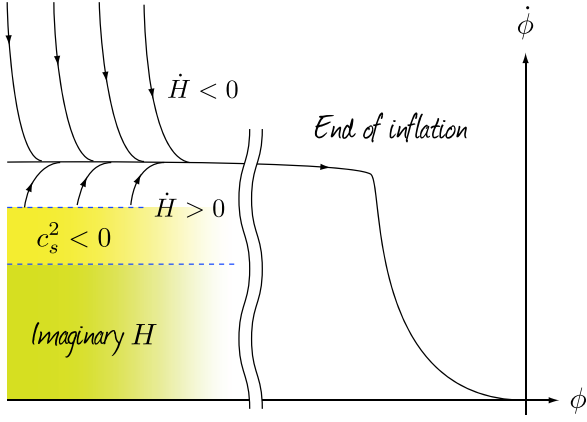}
  \end{center}
  \caption{Schematic phase space diagram of G-inflation.
  The line $\dot H=0$ does not coincide with the line $c_s^2=0$ in general
  so that stable violation of the null energy condition is possible.
  }%
  \label{fig:phase.eps}
\end{figure}

The phase space diagram of G-inflation is depicted in Fig.~\ref{fig:phase.eps}.
Note that in G-inflation
the null energy condition (NEC) may be violated,
{\em i.e.}, $2\mpl^2\dot H=-(\rho+p)>0$.
The NEC violation can occur {\em stably}~\cite{NV, vikman},
in the sense that the squared sound speed (to be defined shortly) is positive
and there are not any ghosts. (This is the condition for the stability
concerning the short wavelength perturbations.)
We stress that this can never occur
in a healthy manner in k-inflation~\cite{kinflation}.
Moreover, NEC violating k-inflation cannot end,
as argued in~\cite{phantom}.

We now move on to study scalar perturbations in this model
using the unitary gauge with $\delta\phi=0$ and
\begin{eqnarray}
\D s^2=-(1+2\alpha)\D t^2+2a^2\partial_i\beta \D t\D x^i
+a^2(1+2{\cal R}_\phi)\D \mathbf{x}^2.
\end{eqnarray}
In this gauge we have $\delta T_i^{\;0}=-G_X\dot\phi^3\partial_i\alpha$,
and hence this gauge does not coincide with
the comoving gauge $\delta T_i^{\;0}=0$.
Consequently, ${\cal R}_\phi$ in general differs from
the comoving curvature perturbation ${\cal R}_c$.
This point highlights the difference between the present model
and the standard k-inflationary model
described simply by ${\cal L}_\phi=K(\phi, X)$~\cite{per-k-inf}.
It will turn out that the variable ${\cal R}_\phi$
is subject to an analogous wave equation to the familiar Sasaki-Mukhanov equation.

Expanding the action~(\ref{action}) to second order in 
the perturbation variables
and then substituting the Hamiltonian and momentum constraint 
equations to eliminate $\alpha$ and $\beta$,
we obtain the following quadratic action for ${\cal R}_\phi$:
\begin{eqnarray}
S^{(2)}=\frac{1}{2}\int\D\tau\D^3x\,
z^2\left[{\cal G}({\cal R}_\phi')^2-{\cal F}(\Vec{\nabla}{\cal R}_\phi)^2\right],
\label{2ndorderaction}
\end{eqnarray}
where
\begin{eqnarray}
z&:=&\frac{a\dot\phi}{H-G_X\dot\phi^3/2\mpl^2},
\\
{\cal F}&:=&K_X+2G_X\left(\ddot\phi+2H\dot\phi\right)
-2\frac{G_X^2}{\mpl^2}X^2
\nonumber\\&&
+2G_{XX}X\ddot\phi-2\left(G_\phi-XG_{\phi X}\right),
\\
{\cal G}&:=&K_X+2XK_{XX}+6G_{X}H\dot\phi+6\frac{G_X^2}{\mpl^2}X^2
\nonumber\\&&
-2\left(G_\phi+XG_{\phi X}\right)+6G_{XX}HX\dot\phi,
\end{eqnarray}
and the prime represents differentiation with respect to the conformal time $\tau$.
The squared sound speed is therefore
$c_s^2={\cal F}/{\cal G}$.
To avoid ghost and gradient instabilities
we require the conditions
${\cal F}>0$ and ${\cal G}>0$.  
One should note that the above equations have been derived without
assuming any specific form of $K(\phi, X)$
and $G(\phi, X)$.


It is now easy to check whether a given G-inflation model is stable or not.
In the simplest class of models (\ref{Ginf}),
we have
\begin{eqnarray}
{\cal F}=-\frac{K_X}{3}+\frac{XK_X^2}{3K},
\;
{\cal G}=-K_X+2XK_{XX}-\frac{XK_X^2}{K},
\end{eqnarray}
where the ``slow-roll'' suppressed terms are ignored.
For the previous toy model~(\ref{toymodel}) one obtains
${\cal F}=x(1-x)/6(1-x/2)$ and ${\cal G}=1-x+(1-x/2)^{-1}$.
Since $0<x<1$, both ${\cal F}$ and ${\cal G}$ are positive. In this model,
the sound speed is smaller than the speed of light: $c_s^2\le (4\sqrt{2}-5)/21\simeq 0.031<1$.

In the superhorizon regime where ${\cal O}(\Vec{\nabla}^2)$ terms can be 
neglected,
the two independent solutions to the perturbation equation
that follows from the action~(\ref{2ndorderaction})
are
\begin{eqnarray}
{\cal R}_\phi=\text{const},\quad
\int^\tau\!\frac{\D\tau'}{z^2{\cal G}}.
\end{eqnarray}
The latter is a decaying mode in the inflationary stage and in the
subsequent reheating stage in our model, and hence can be neglected.
In this limit one can show that ${\cal R}_\phi$ coincides with
the comoving curvature perturbation.



The power spectrum of
${\cal R}_\phi$ generated during G-inflation can be evaluated 
by writing the perturbation equation
in the Fourier space as
\begin{eqnarray}
\frac{\D^2u_k}{\D y^2}+\left(k^2-\frac{\tilde z_{,yy}}{\tilde z}\right)u_k=0,
\end{eqnarray}
where
$\D y=c_s\,\D \tau$, $\tilde z:=\left({\cal F}{\cal G}\right)^{1/4}z$, and
$u_k:=\tilde z{\cal R}_{\phi,k}$.
Let us again focus on the class of models~(\ref{Ginf}).
Note that
the sound speed $c_s$ may vary rapidly in the present case,
and hence one cannot neglect $\epsilon_s:=\dot c_s/Hc_s$
even when working in leading order in ``slow-roll.''
Indeed, one finds
$\epsilon_s\simeq \eta X ( {\cal G}_X/{\cal G} - {\cal F}_X/{\cal F} )$.
With some manipulation, one obtains
$\tilde z_{,yy}/\tilde z \simeq (-y)^{-2}[2+3\epsilon{\cal C}(X)]$ with
\begin{eqnarray}
{\cal C}(X):=\frac{K}{K_X}\frac{Q_X}{Q},
\quad
Q(X):=\frac{(K-XK_X)^2}{18\mpl^4 Xc_s^2\sqrt{{\cal F}{\cal G}}}.
\end{eqnarray}
It should be emphasized that scalar fluctuations are generated
even from exactly de Sitter inflation. This is because, as mentioned
before, the Galileon symmetry is broken in the de Sitter background,
which is manifest from $\dot\phi=$ const. This situation is in
stark contrast with other inflation models: scalar fluctuations cannot
be generated from the de Sitter background with $\dot\phi=0$ in usual
potential-driven inflation, while the exactly de Sitter background cannot
be realized in k-inflation.

The normalized mode is given in terms of the Hankel function as
\begin{eqnarray}
u_k=\frac{\sqrt{\pi}}{2}\sqrt{-y}H_\nu^{(1)}(-k y),
\quad \nu:=\frac{3}{2}+\epsilon\,{\cal C},
\end{eqnarray}
from which it is straightforward to obtain the
power spectrum and the spectral index:
\begin{eqnarray}
{\cal P}_{{\cal R}_\phi}=\left.\frac{Q}{4\pi^2}\right|_{c_s k=1/(-\tau)},
\quad
n_s-1=-2\epsilon\,{\cal C}.\label{sind}
\end{eqnarray}

The behavior of tensor perturbations in G-inflation is basically the
same as in the usual inflation models and is completely determined
geometrically.  Therefore, the power spectrum and the spectral index of
primordial gravitational waves are given by ${\cal
P}_T=(8/\mpl^2)(H/2\pi)^2$ and $n_T=-2\epsilon$. However, it would
be interesting to point out that the tensor spectrum can be blue in
G-inflation with possible violation of the NEC.
The positive tensor spectral index not only is compatible with
current observational data,
but also broadens the limits on cosmological parameters~\cite{blue}.
Moreover, the amplitude of tensor fluctuation with such a blue spectral
index is relatively enhanced for large frequencies, which makes its
direct detection easier.


As a concrete example, let us come back again to the previous toy
model~(\ref{toymodel}), in which the tensor-to-scalar ratio is given by
\begin{eqnarray}
r\simeq \frac{16\sqrt{6}}{3}\left(\frac{\sqrt{3}\mu}{\mpl}\right)^{3/2}
\quad\text{for}\quad \mu\ll \mpl.
\end{eqnarray}
With the properly normalized scalar perturbation, ${\cal P}_{{\cal
R}_\phi} = 2.4\times 10^{-9}$, we can easily realize large $r$ to
saturate the current observational bound, exceeding the predictions of
the chaotic inflation models \cite{chaotic}.\footnote{Another interesting
inflation model with the enhanced tensor-to-scalar ratio has been
proposed in~\cite{mv}, which relies on a sound speed greater than the speed of light.}
For example, for
$M=0.00425\times \mpl$ and $\mu=0.032\times \mpl$ we find $r=0.17$,
which is large enough to be probed by the PLANCK satellite
\cite{PLANCK}.
Note that neither the standard consistency relation, $r=-8n_T$,
nor the k-inflation-type consistency relation, $r=-8c_sn_T$, holds in our model.

In summary, we have proposed a novel inflationary mechanism driven by
the Galileon-like scalar field.  Our model ---{\em G-inflation}--- is a
new class of inflation models with the term proportional to $\Box\phi$
in the Lagrangian, which opens a new branch of inflation model building.
Contrary to the most naive expectation, the interaction of the form
$G\left(\phi, (\nabla\phi)^2\right)\Box\phi$ gives rise to derivatives
no higher than two in the field equations~\cite{vikman}.  In this sense, G-inflation
is distinct also from ghost condensation~\cite{ghost} and
B-inflation~\cite{binflation}. After G-inflation, the universe is
reheated through the gravitational particle production with successful
thermal leptogenesis.  We have also shown that G-inflation can generate
(almost) scale-invariant density perturbations, possibly together with a
large amplitude of primordial gravitational waves.
These facts have great impacts on the planned and ongoing gravitational
wave experiments and CMB observations. In a forthcoming paper we shall
compute the non-Gaussianity of the curvature perturbation from
G-inflation~\cite{toappear}, which would be a powerful discriminant of
the scenario in addition to the violation of the standard consistency
relation.

We would like to thank Takehiko Asaka for useful comments. This work was
supported in part by JSPS Grant-in-Aid for Scientific Research
Nos. 19340054 (J.Y.), 21740187 (M.Y.), and the Grant-in-Aid for
Scientific Research on Innovative Areas No. 21111006 (J.Y.).

{\bf Note added~~} While this paper was being completed,
Ref.~\cite{vikman} appeared, in which the quadratic action for
cosmological perturbations is given independently, though it is
investigated in a different context, that is, dark energy.



\end{document}